# Elastic properties of mono- and polycrystalline hexagonal $AlB_2$ – like diborides of *s, p* and *d* metals from first-principles calculations

Igor R. Shein[a)] and Alexander L. Ivanovskii

*Institute of Solid State Chemistry, Ural Branch of the Russian Academy of Sciences, 620041 Ekaterinburg, Russia*

We have performed accurate *ab initio* total energy calculations using the full-potential linearized augmented plane wave (FP-LAPW) method with the generalized gradient approximation (GGA) for the exchange-correlation potential to systematically investigate elastic properties of 18 stable, meta-stable and hypothetical hexagonal ($AlB_2$ – like) metal diborides $MB_2$, where M = Na, Be, Mg, Ca, Al, Sc, Y, Ti, Zr, Hf, V, Nb, Ta, Cr, Mo, W, Ag and Au. For monocrystalline $MB_2$ the optimized lattice parameters, independent elastic constants ($C_{ij}$), bulk modules ($B$), shear modules ($G$) are obtained and analyzed in comparison with the available theoretical and experimental data. For the first time numerical estimates of a set of elastic parameters of the polycrystalline $MB_2$ ceramics (in the framework of the Voigt-Reuss-Hill approximation), namely bulk and shear modules, compressibility ($\beta$), Young's modules ($Y$), Poisson's ratio ($\nu$), Lamé's coefficients ($\mu, \lambda$), are performed.

*Keywords:* First principle; Metal diborides; Elastic properties, Polycrystalline ceramic

---

[a)]Electronic mail: shein@ihim.uran.ru

## I. INTRODUCTION

Transition metal diborides $MB_2$, as well as $MB_2$-based solid solutions and composites, attract for a long time much attention of physicists and material scientists due to their unique physical and chemical properties such as hardness, high melting point, chemical inertness, etc., and belong to the most promising engineering materials with a wide range of industrial applications. [1-12] Besides bulk $MB_2$ materials, increasing attention is devoted recently to their nano-sized forms such as nanopowders, nanowires, nanotubes and nanocomposites. [13-20]

An unexpected discovery [21] of superconductivity with the temperature of critical transition $T_C \sim 39K$ in magnesium diboride $MgB_2$ has renewed a new powerful stimulus to studying this family of materials, see reviews. [22-31] Additionally, the interest in metal diborides (such as $OsB_2$, $ReB_2$, $RuB_2$ [32-38]) is increased now due to search of new ultra-incompressible superhard materials.

Besides the above-mentioned exciting physical properties, metal diborides also adopt several crystal structures depending on the $R_M/R_B$ ratio, where $R_M$ and $R_B$ are atomic radii of metallic and boron atoms, respectively. The most representative group of $MB_2$ borides is formed by the layered hexagonal diborides with $AlB_2$ - like structural type. For these stable $AlB_2$ - like $MB_2$ phases $R_M/R_B$ ratio varies in range from 1.14 to 2.06.[39]

Along with the above $MgB_2$ and $AlB_2$, the most known $MB_2$ phases are diborides of transition metals of IV-VI groups (Ti, Zr, Hf, Nb etc.). With filling of the TM *d* shell, the stability of these phases rapidly decreases, and $MoB_2$ and $WB_2$ lie on the border of stability. For example, the experimental data on Mo-B phase equilibria [5,7,40-48] are quite contradictory: in the composition region near to "molybdenum diboride" the presence of hexagonal $MoB_2$, rhombohedral $Mo_2B_5$, as well as the metal-deficient ($Mo_{1-x}B_2$ or $Mo_{2-x}B_5$) and boron-deficient compositions ($MoB_{2-x}$ and also $Mo_2B_{5-x}$) is reported depending on the synthetic route and the thermodynamic parameters of the process. So, the products obtained [47] by mechano-chemical synthesis without external heat ($MoB_2$, $Mo_2B_5$ or $MoB_2$ +

$Mo_2B_5$) are strongly determined by the relation of co-milled boron and Mo powder mixtures. On the contrary, upon boronizing of Mo in molten salts, [48] only the $Mo_2B_5$ phase was obtained.

In addition, a set of hypothetical $AlB_2$-like $MB_2$ phases (M = Li, Na, Be, Ca, Sr, Ag and Au) was discussed recently, and their properties were analyzed by means of theoretical approaches. [26-28,49-56] So, some theoretical efforts were made to predict the superconducting transition temperature $T_C$ and the stability of these materials. [26-28,49-56] Let's note that despite some experimental efforts, [57,58] undoped samples of these species have not been prepared up to now .

Despite a great number of interesting results on physical properties for $AlB_2$ -like diborides, which were reported in past years, their mechanical behavior is sufficiently investigated only for several phases: $MgB_2$, $TiB_2$ and $ZrB_2$, whereas for the majority of other diborides these data, to our knowledge, are rather limited. Besides, most of the theoretical works devoted to the elastic behavior of $MB_2$ are performed by means of different approximations and, moreover, deal exclusively with monocrystals. Meanwhile, the most part of synthesized and experimentally examined $MB_2$ materials are polycrystalline.

In this paper, in order to get systematic insight into the elastic properties of these materials, a comparative first-principle study for 18 $AlB_2$-like diborides of I-VI groups *s, p* and *d* metals (among them there are stable: $MgB_2$, $AlB_2$, $ScB_2$, $YB_2$, $TiB_2$, $ZrB_2$, $HfB_2$, $VB_2$, $NbB_2$, $TaB_2$ and $CrB_2$, as well as meta-stable: $MoB_2$ and $WB_2$, and hypothetical phases: $NaB_2$, $BeB_2$, $CaB_2$, $AgB_2$ and $AuB_2$) has been performed using the FLAPW method within the generalized gradient approximation (GGA) for the exchange-correlation potential.

As a result, we have evaluated and analyzed a set of physical parameters of the above-mentioned $MB_2$ monocrystals such as optimized lattice parameters, density, elastic constants ($C_{ij}$), bulk modules ($B$) and shear modules ($G$). Additionally, the numerical estimates of mechanical parameters: elastic modules ($B$, $G$), compressibility ($\beta$), Young's modules ($Y$), Poisson's ratio ($\nu$), Lamé's coefficients ($\mu$, $\lambda$) for the corresponding polycrystalline $MB_2$ ceramics (in the Voigt-Reuss-Hill approximation) were obtained and analyzed for the first time.

## II. METHOD AND DETAILS OF CALCULATIONS

The $AlB_2$-like metal diborides adopt the layered crystal structure (space group P6/*mmm*). It is a simple hexagonal lattice of close-packed metal layers alternating with graphite-like B layers in sequence ..*AHAHAH*.. perpendicularly to the *c* direction.[1-6] The boron atoms are arranged at the corners of a hexagon with three nearest neighbor B atoms in each plane. The metal atoms are located at the center of the B hexagons, midway between adjacent boron layers. Thus the M and B atoms have the [$MB_{12}M_8$] and [$BM_6B_3$] coordination polyhedra, respectively. Each metal center has $D_{6h}$ symmetry, *i.e.* has 12 borons at the vertices of a hexagonal prism. In addition, the central metal atom is coordinated also by 8 metals through the faces of the $B_{12}$ prisms. The primitive cell contains 1 metal and 2 boron atoms (Z = 1). There is one B and one metal in nonequivalent atomic positions of M (0, 0, 0) and B (1/3, 2/3, 1/2).

The calculations of the all mentioned diborides (Table I) were carried out by means of the full-potential method with mixed basis APW+lo (LAPW) implemented in the WIEN2k suite of programs.[59] The generalized gradient correction (GGA) to exchange-correlation potential of Perdew, Burke and Ernzerhof [60] was used. The maximum value for partial waves used inside atomic spheres was $l$ = 12 and the maximum value for partial waves used in the computation of muffin-tin matrix elements was $l$ = 4. The plane-wave expansion with $R_{MT} \times K_{MAX}$ equal to 7, and $k$ sampling with 10×10×10 $k$-points mesh in the Brillouin zone was used. Relativistic effects were taken into account within the scalar-relativistic approximation. The self-consistent calculations were considered to converge when the difference in the total energy of the crystal did not exceed 0.001 mRy as calculated at consecutive steps. Other details of the calculations are described in Ref. 46.

## III. RESULTS AND DISCUSSION

### 3.1. Monocrystalline $MB_2$ phases

Firstly, the equilibrium lattice constants (*a* and *c*) for all $MB_2$ phases are obtained from the total energy calculations, followed by the fitting of these results to the Birch-Murnaghan equation. [61] The results obtained are in reasonable agreement with the available theoretical and experimental data, Table II. Let's note that in Table II the calculated ratio *c*/*a* for each $MB_2$ phase is also presented. According to Pearson's criterion, [79] for stable $AlB_2$-like phases (see also below) the ratio *c*/*a* should lie in the interval from 0.59 to 1.2. Thus, from the data obtained it may be concluded that the diborides $NaB_2$, $CaB_2$, $AgB_2$ and $AuB_2$, for which $c/a > 1.2$, should be unstable, whereas $MgB_2$ ($c/a$ =1.15) and $YB_2$ ($c/a$ = 1.18) are near to the upper border of stability.

Let us discuss the mechanical parameters as obtained within the framework of the FLAPW–GGA calculations for $MB_2$ monocrystals. The values of five independent elastic constants ($C_{ij,}$ namely $C_{11}$, $C_{12}$, $C_{13}$, $C_{33}$ and $C_{44}$; while $C_{66} = ½ (C_{11} - C_{12})$) for these phases summarized in Table III were found by imposing five different deformations (monoclinic, triclinic and three - of hexagonal types, see Ref. 80) to the equilibrium lattice of the hexagonal unit cell and by determining the dependence of the resulting energy changes on the deformation. There are some general conclusions we would like to point out from the obtained results.

Firstly, mechanically stable phases should satisfy the well-known Born criteria: $C_{11} >0$, $(C_{11} - C_{12}) >0$, $C_{44} >0$, and $(C_{11} + C_{12})C_{33} - 2C_{12}^2 > 0$. In our case, these conditions are not met for four phases: $BeB_2$, $AgB_2$ and $AuB_2$, for which $C_{44} < 0$, and for $NaB_2$: $(C_{11} - C_{12}) < 0$; additionally, for all the four diborides $(C_{11} + C_{12})C_{33} - 2C_{12}^2 < 0$. This means that these hypothetical phases are mechanically unstable. This result agrees with theoretical estimations of heat of formations for $NaB_2$, $BeB_2$, $AgB_2$ and $AuB_2$,[53] according to which these phases should be energetically unstable - in comparison with the mechanical mixture of the constituent reagents. Other arguments about instability of these phases, in terms of electronic density of states and chemical bonding picture, are discussed in Refs. 26,28,51-55. Thus, these phases will not be analyzed below.

Secondly, the estimations of the anisotropy factor $A = C_{33}/C_{11}$ (Table III) demonstrate larger elastic anisotropy in the mechanical properties for *s* and *p* metal diborides, as well as for meta-stable $MoB_2$ and $WB_2$ - in comparison with moderate anisotropy for other *d* metal diborides. Probably, this tendency may originate from strong M *d* – B *p* bonding between metal and boron sheets in these compounds - in comparison with weaker *p-p* and *s-p* bonding for *s* and *p* metal diborides, whereas for $MoB_2$ and $WB_2$ the reduction of intra-layer interactions (i.e. the growth of anisotropy) will be the result of filling of anti-bonding *d* bands. [26,28,51-55] Besides, for anisotropic systems the large $C_{13}/C_{12}$ and small $C_{33}/C_{11}$ values indicate that atomic bonding along the *x* axis is stronger than that along the *z* axis. In our case in the sequence $MB_2$ (where M are *s* and *p* metals) → $MB_2$ (where M are *d* metals) the values of $C_{13}/C_{12}$ drastically increase (for example from 0.52 ($MgB_2$) and 0.82 ($AlB_2$) to 1.54 ($TiB_2$) and 1.85 ($NbB_2$)), whereas $C_{33}/C_{11}$ values remain comparable enough (for example: 0.65 ($MgB_2$), 0.51 ($AlB_2$), 0.71 ($TiB_2$) and 0.81 ($NbB_2$)). This fact may be explained as strengthening of metal-metal bonds for *d* metal diborides.

The calculated elastic constants allow us to obtain the macroscopic mechanical parameters of $MB_2$ phases, namely their bulk ($B$) and shear ($G$)modules. The isotropic bulk modules $B_{iso}$ are obtained under the assumption that the $c/a$ ratio does not change when the lattice is subjected to an isotropic stress. The relevant expression for $B_{iso}$ [91] coincides with the definition of the bulk modules in Voigt (V) [92] approximation:

$$B_V = \frac{2}{9}\left(C_{11} + C_{12} + 2C_{13} + \frac{1}{2}C_{33}\right)$$

In the same approximation shear modules ($G_V$) are estimated as:

$$G_V = \frac{1}{30}(C_{11} + C_{12} + 2C_{33} - 4C_{13} + 12C_{55} + 12C_{66})$$

The results obtained are summarized in Table IV in comparison with other theoretical data. From our results we see that both bulk and shear modules for *s* and *p* metal diborides are as a whole lower than for *d* metal diborides. It is usually assumed that hardness of the materials is defined by these elastic modules: the bulk modulus is a measure of resistance to volume change by applied pressure, whereas the shear modulus is a measure of resistance to reversible deformations upon shear stress. [93] Thus, the *d* metal diborides are the hardest materials among all $MB_2$ phases.

On the other hand, according to criterion,[94] a material is brittle if the *B/G* ratio is less than 1.75. In our case, for all the diborides, except $WB_2$ (for which B/G ~ 1.9, Table IV), these parameters are much less than 1.75, *i.e.* these materials will behave in a brittle manner.

Finally, quite interesting conclusions follow from comparison of our data for $MB_2$ phases and elastic properties of other binary phases, in particular, *d* metal monocarbides (MC) and mononitrides (MN), see Fig. 1. Comparison of our bulk modules for $MB_2$ with the results [95] for MC and MN shows that the general trends in the dependence of *B* on the *d* metal type are very similar to each other. Namely: (i) the bulk modules of all binary phases for III group metals are minimal; (ii) the bulk modules of binary phases for III and IV group metals are considerably different, whereas the bulk modules of phases for V and VI group metals are more close to each other; (iii) the bulk modules increase for all binary phases of 3*d* - 5*d* metals as going from III to VI group; and (iv) in the same groups the bulk modules decrease as going from 3*d* to 4*d* metals and increase as going from 4*d* to 5*d* metals. One noticeable difference is seen for Ta and W – based phases, where $B(TaB_2) > B(WB_2)$, whereas for the corresponding carbides and nitrides $B$(Ta(C,N)) < $B$(W(C,N)). Obviously, this disagreement should be ascribed to the above meta-stability of the $AlB_2$-like tungsten diboride in the $AlB_2$-like structure because extra electrons fill the antibonding states.[26,53] The mentioned similarities in the change of the bulk modules of $MB_2$, MC and MN phases in the dependence from the *d* metal type may seem quite surprising taking into account the fact that $MB_2$ and MC, MN phases adopt (i) very different structures (layered $AlB_2$ type *versus* isotropic *B*1 type) and (ii) different nature of inter-atomic bonding (strong covalent B-B intra-layer interactions plus covalent B-M intra-layer interactions for $MB_2$ [26,28,51-55] *versus* dominant covalent M-(C,N) bonding for MC and MN [95-99]). Therefore we may suppose that the trends in elastic moduli changes for all these phases are controlled mainly by their cell volumes (*i.e.* atomic radii of *d* metals, and follow the known semiempirical expression [100-102] $B \sim 1/l^n$, where *l* is the bond length) and depend to a much lesser degree on the differences in structure, cohesion energies and the types of bonds in these phases, see also below.

### 3.2. Polycrystalline $MB_2$ ceramics

The above elastic parameters are obtained from first-principle calculations of $MB_2$ monocrystals. Meanwhile, the majority of synthesized and experimentally examined $MB_2$ materials are prepared and investigated as polycrystalline ceramics, [1-7] *i.e.* in the form of aggregated mixtures of microcrystallites with a random orientation. Thus, it is useful to estimate the elastic parameters for the polycrystalline $MB_2$ materials.

For this purpose we utilize the Voigt-Reuss-Hill (VRH) approximation. In this approach, according to Hill, [103] two main approximations: Voigt [92] and Reuss (R) [104] are used. In turn, in Reuss's model the bulk and shear modules ($B_R$ and $G_R$) are estimated as:

$$B_R = \frac{(C_{11}+C_{12})C_{33}-2C_{12}^2}{C_{11}+C_{12}+2C_{33}-4C_{13}}$$

$$G_R = \frac{5}{2}\frac{\left[(C_{11}+C_{12})C_{33}-2C_{12}^2\right]C_{55}C_{66}}{3B_V C_{55}C_{66}+\left[(C_{11}+C_{12})C_{33}-2C_{12}^2\right](C_{55}+C_{66})}$$

Then, the Voigt and Reuss averages are limits and the actual effective modules for polycrystals could be estimated in VRH approach by the arithmetic mean of these two bounds. In this way, when the bulk modulus ($B_{VRH}$) and the shear modulus ($G_{VRH}$) are obtained from $B_{V,R}$ and $G_{V,R}$ by the VRH approach in simple forms as: $B_{VRH}$ = ½ ($B_V$ + $B_R$) and $G_{VRH}$ = ½ ($G_V$ + $G_R$), averaged compressibility ($\beta_{VRH}$ = 1/$B_{VRH}$) and Young's modules ($Y_{VRH}$) may be calculated by the expression:

$$Y_{VRH} = \frac{9B_{VRH}G_{VRH}}{3B_{VRH} + G_{VRH}}$$

Then, the Poisson's ratio ($\nu$) and Lame's constants ($\mu, \lambda$) were obtained for polycrystalline $MB_2$ species from $B_{VRH}$, $G_{VRH}$ and $Y_{VRH}$ as:

$$\nu = \frac{3B_{VRH} - 2G_{VRH}}{2(3B_{VRH} + G_{VRH})}$$

$$\mu = \frac{Y_{VRH}}{2(1+\nu)}; \quad \lambda = \frac{\nu Y_{VRH}}{(1+\nu)(1-2\nu)}$$

The above-mentioned parameters are summarized in Table V in comparison with available experimental data for $MB_2$ ceramics. Here we should point out that our estimations are made in the limit of zero porosity of $MB_2$ ceramics.

It was found for the majority of $MB_2$ ceramics that $B_{VRH} > G_{VRH}$; this implies that the parameter limiting the stability of $MB_2$ materials is the shear modulus $G_{VRH}$. We have obtained only two deviations from this tendency: $ScB_2$ and $TiB_2$, for which $B_{VRH} \sim G_{VRH}$.

The trends of bulk modules for $MB_2$ ceramics may be discussed as displayed in Fig. 2. As can be seen, the bulk modules increase as the *d* metal goes from the III group to the VI group, or downward on the Periodic Table; the borides of *d* metals have larger $B_{VRH}$ than their *s*, *p* metal counterparts. As is seen, there is the above-mentioned distinct correlation between elastic modules and unit cell volumes of $MB_2$ phases. Generally, these trends are supported by experiment, Table V.

Thus, diborides of *s*, *p* metals, $ScB_2$ and $YB_2$ adopt the minimal $B_{VRH}$, whereas the bulk modules for diborides of *d* metals of IV- VI groups can be varied from 238 to 437 GPa. The maximal bulk modulus (437 GPa) and the minimal compressibility (0.00229 1/GPa) have been obtained for $WB_2$ ceramic. On the other hand, among IV- VI group *d*-metal diborides, $TiB_2$, $HfB_2$, $VB_2$ and $TaB_2$ will have the maximal shear modules ($G_{VRH}$ ~ 240 - 270 GPa), *i.e.* the maximal bond-restoring energy under elastic shear strain , whereas, according to the $G_{VRH}$ estimations, $CrB_2$ ($G_{VRH}$ = 174 GPa) will remain a material with the minimal hardness. The Young's modulus $Y_{VRH}$ has also the maximal and minimal values for $TiB_2$, $HfB_2$, $VB_2$ and $TaB_2$ ($Y_{VRH}$ ~ 560-600 GPa) and $CrB_2$ ceramics ($Y_{VRH}$ = 442 GPa), respectively.

The values of the Poisson ratio ($v$) for covalent materials are small ($v$ ~ 0.1), whereas for metallic materials $v$ is typically 0.33.[105] In our case the values of $v$ for $MB_2$ vary from about 0.120 - 0.166 for $(Ti,Zr,Hf)B_2$ to 0.267-0.310 for $(Cr,Mo,W)B_2$, indicating an increase of metal-metal bonding for the diborides when *d* metal goes from the IV group to the VI group. This trend reflects the well-known situation [26,28,53-55] when the metallicity of $MB_2$ phases increases with the filling of *d* orbitals.

## IV. CONCLUSIONS

In summary, we have performed FLAPW-GGA calculations to obtain the systematic trends for elastic properties of 18 hexagonal $AlB_2$-like diborides of *s*, *p* and *d* metals, among them there are stable ($MgB_2$, $AlB_2$, $ScB_2$, $YB_2$, $TiB_2$, $ZrB_2$, $HfB_2$, $VB_2$, $NbB_2$, $TaB_2$ and $CrB_2$), meta-stable ($MoB_2$ and $WB_2$) and hypothetical phases ($NaB_2$, $BeB_2$, $CaB_2$, $AgB_2$ and $AuB_2$). For the first time a set of elastic parameters of the polycrystalline $MB_2$ ceramics has been estimated.

Our analysis of elastic constants ($C_{ij}$) shows that $NaB_2$, $BeB_2$, $AgB_2$ and $AuB_2$ belong to mechanically unstable systems. The present study shows also that both bulk and shear modules for *s* and *p* metal diborides are as a whole lower than for *d* metal diborides, *i.e.* *d* metal diborides should be the hardest materials among all the examined $MB_2$ phases. On the other hand, all diborides, except $WB_2$, will behave in a brittle manner.

From our studies of $MB_2$ ceramics we may draw the following main conclusions:

(i) For the majority of diborides $B_{VRH} > G_{VRH}$; this implies that a parameter limiting the stability of $MB_2$ materials is the shear modulus $G_{VRH}$;

(ii) The values of the Poisson ratio vary from about 0.120 - 0.166 for $(Ti,Zr,Hf)B_2$ to 0.267-0.310 for $(Cr,Mo,W)B_2$, indicating an increase of metal-metal bonding for the diborides when *d* metal goes from the IV group to the VI group. This trend can be understood in terms of filling of the relevant d-like electronic bands.

(iii) The bulk moduli increase as the *d* metal goes from the III group to the VI group, or downward on the Periodic Table; the borides of *d* metals have larger $B_{VRH}$ than their *s*, *p* metal counterparts. These trends correlate with the cell volumes of $MB_2$ and generally are supported by experiment.

## ACKNOWLEDGEMENTS

The work was supported by RFBR, grant 08-08-00034-a.

**TABLE I.** (*Color online*). Considered hexagonal ($AlB_2$-like) diborides of *s*, *p* and *d* metals. *

| | I | II | III | IV | V | VI |
|---|---|---|---|---|---|---|
| 2 | | Be | | | | |
| 3 | Na | Mg | Al | | | |
| 4 | | Ca | Sc | Ti | V | Cr |
| 5 | Ag | | Y | Zr | Nb | Mo |
| 6 | Au | | | Hf | Ta | W |

1 2 3

* 1-stable, 2- meta-stable and 3 – hypothetical $MB_2$ phases.

**TABLE II**. Calculated lattice constants (*a, c*, in Å) for hexagonal ($AlB_2$-like) diborides of *s, p* and *d* metals in comparison with available experimental and theoretical data.

| diboride | *a* | *c* | *c/a* |
|---|---|---|---|
| **$NaB_2$** | **3.002** | **4.297** | **1.43** |
| | 2.995 [a] | 4.229 [a] | 1.41 [a] |
| **$BeB_2$** | **2.898** | **2.884** | **0.99** |
| | 2.896, [a] 2.886 [b] | 2.845, [a] 3.088 [b] | 0.98, [a] 1.06 [b] |
| **$MgB_2$** | **3.050** | **3.511** | **1.15** |
| | 3.040,[a] 3.080, [b] 3.085, [c]<br>3.098, [d] 3.073 [e] 3.071 [f]<br>3.064, [g] 3.049, [q] | 3.448, [a] 3.532, [b] 3.523 [c]<br>3.520, [d] 3.527, [e] 3.528 [f]<br>3.493, [g] 3.446, [q] | 1.13, [a] 1.15, [b] 1.14, [c]<br>1.14, [d] 1.15, [e] 1.15 [f]<br>1.14, [g] 1.13 [q] |
| **$CaB_2$** | **3.191** | **4.051** | **1.27** |
| | 3.183, [a] 3.397, [b] 3.205 [h] | 4.001, [a] 4.019, [b] 3.974 [h] | 1.26, [a] 1.18, [b] 1.24 [h] |
| **$AlB_2$** | **2.962** | **3.206** | **1.08** |
| | 2.978, [a] 3.005, [c] 3.008, [l]<br>3.009, [m] 2.998, [n] | 3.248, [a] 3.257, [c] 3.261, [l]<br>3.262, [m] 3.286, [n] | 1.09, [a] 1.08, [c] 1.08, [i]<br>1.08, [m] 1.10 [n] |
| **$ScB_2$** | **3.114** | **3.512** | **1.13** |
| | 3.114, [a] 3.148, [c] | 3.465, [a] 3.517, [c] | 1.11, [a] 1.12 [c] |
| **$YB_2$** | **3.268** | **3.849** | **1.18** |
| | 3.253, [a] 3.290, [c] | 3.812, [a] 3.835, [c] | 1.17, [a] 1.16 [c] |
| **$TiB_2$** | **3.006** | **3.212** | **1.07** |
| | 3.005, [a] 3.038, [c] 3.070, [b]<br>3.029, [e] 3.027, [i] 2.993 [j]<br>3.015, [k] | 3.186, [a] 3.239, [c] 3.262, [b]<br>3.220, [e] 3.240, [i] 3.147, [j]<br>3.222, [k] | 1.06, [a] 1.07, [c] 1.06, [b]<br>1.06, [e] 1.07, [i] 1.05, [j]<br>1.07, [k] |
| **$ZrB_2$** | **3.155** | **3.542** | **1.12** |
| | 3.139, [a] 3.130, [c] 3.127, [o]<br>3.183, [p] 3.169, [q] | 3.499, [a] 3.533, [c] 3.523, [o]<br>3.546, [p] 3.531, [q] | 1.12, [a] 1.13, [c] 1.13, [o]<br>1.11, [p] 1.11, [q] |
| **$HfB_2$** | **3.144** | **3.502** | **1.11** |
| | 3.111, [a] 3.141, [c] | 3.409, [a] 3.470, [c] | 1.10, [a] 1.11, [c] |
| **$VB_2$** | **2.970** | **3.029** | **1.02** |
| | 2.969, [a] 2.998, [c] 3.008, [p]<br>3.007, [q] | 2.989, [a] 3.005, [c] 3.068, [p]<br>3.048, [q] | 1.01, [a] 1.00, [c] 1.02, [p]<br>1.01, [q] |
| **$NbB_2$** | **3.086** | **3.318** | **1.08** |
| | 3.086, [a] 3.086, [c] 3.111, [r]<br>3.181, [q] | 3.304, [a] 3.306, [c] 3.309, [r]<br>3.357, [q] | 1.07, [a] 1.07, [c] 1.06, [r]<br>1.06, [q] |
| **$TaB_2$** | **3.074** | **3.209** | **1.05** |
| | 3.072, [a] 3.088, [c] 3.164, [q] | 3.275, [a] 3.241, [c] 3.323, [q] | 1.07, [a] 1.05, [c] 1.05, [q] |
| **$CrB_2$** | **2.945** | **2.941** | **0.99** |
| | 2.970, [a] 2.973, [c] | 2.880, [a] 3.072, [c] | 0.97, [a] 1.03, [c] |
| **$MoB_2$** | **3.009** | **3.305** | **1.10** |
| | 2.990, [a] 3.039, [c] 3.012, [r] | 3.342, [a] 3.055, [c] 3.330, [r] | 1.12, [a] 1.05, [c] 1.11, [r] |
| **$WB_2$** | **2.998** | **3.334** | **1.11** |
| | 2.972, [a] 3.020, [c] 3.053, [s] | 3.369, [a] 3.050, [c] 3.306, [s] | 1.13, [a] 1.01, [c] 1.08, [s] |
| **$AgB_2$** | **2.986** | **4.043** | **1.35** |
| | 2.994, [a] 3.000, [t] 3.034 [v] | 4.051, [a] 3.493, [t] 4.085, [v] | 1.35, [a] 1.16, [t] 1.35, [v] |
| **$AuB_2$** | **2.972** | **4.284** | **1.44** |
| | 2.982, [a] 2.957, [t] | 4.197, [a] 3,785, [t] | 1.41, [a] 1.28 [t] |

[a]Reference 53, FP-LAPW + LDA.
[b]Reference 54, TB-LMTO.

[c]Reference 62, experiment.
[d]Reference 63, FP-LAPW + GGA.
[e]Reference 64, first-principles pseudopotential method.
[f]Reference 66, FLAPW.
[g]Reference 66, DFT (CRYSTAL98).
[h]Reference 51, FP-LAPW + GGA.
[i]Reference 67, Hartree-Fock.
[j]Reference 68, NCP-GGA.
[k]Reference 69, CASTEP-GGA.
[l]Reference 70, plane waves – DFT.
[m]Reference 71, FP-LAPW-GGA.
[n]Reference 72, VASP-GGA.
[o]Reference 73 DFT and GGA.
[p]Reference 74, LCAO-DFT.
[q]Reference 75, FP-LAPW-GGA.
[r]Reference 46, FP-LMTO.
[s]Reference 76, CASTEP-LDA.
[t]Reference 77, LCAO-DFT.
[v]Reference 78, VASP.

**TABLE III**. Calculated elastic constants ($C_{ij}$, in GPa) and anisotropy factors ($A = C_{33}/C_{11}$) for hexagonal ($AlB_2$-like) diborides of *s, p* and *d* metals in comparison with available data.

| diboride | $C_{11}$ | $C_{12}$ | $C_{13}$ | $C_{33}$ | $C_{44}$ | $A$ |
|---|---|---|---|---|---|---|
| **$NaB_2$** | **158.5** | **196.2** | **9.3** | **151.5** | **41.1** | ***u* *** |
| **$BeB_2$** | **358.1**<br>696 [a] | **255.2**<br>136 [a] | **114.4**<br>88 [a] | **106.1**<br>370[a] | **-62.2**<br>119 [a] | ***u*** |
| **$MgB_2$** | **434.8**<br>438, [b] 462, [c]<br>524, [d] 431, [e]<br>446, [f] 365, [g] | **62.1**<br>43,[b] 67, [c]<br>58, [d] 27, [e]<br>68, [f] 98, [g] | **32.8**<br>33, [b] 41,[c]<br>33, [d] 31, [e]<br>39, [f] 65, [g] | **285.4**<br>264, [b] 254,[c]<br>243, [d] 197, [e]<br>284, [f] 203, [g] | **64.4**<br>80,[b] 80, [c]<br>85, [d] 85, [e]<br>77, [f] 58, [g] | **0.65** |
| **$CaB_2$** | **237.5** | **97.8** | **34.9** | **289.8** | **96.9** | **1.22** |
| **$AlB_2$** | **530.0**<br>665 [h] | **82.4**<br>41 [h] | **67.9**<br>17 [h] | **272.3**<br>417 [h] | **32.7**<br>58 [h] | **0.51** |
| **$ScB_2$** | **508.5** | **41.9** | **73.5** | **373.3** | **189.5** | **0.73** |
| **$YB_2$** | **353.9** | **56.6** | **85.9** | **326.7** | **166.3** | **0.92** |
| **$TiB_2$** | **670.9**<br>656, [g] 786, [i]<br>588-711, [j]<br>654, [k] 671,[l]<br>626 [m] | **64.0**<br>66,[g] 127, [i]<br>17-410, [j]<br>75,[k] 62, [l]<br>68 [m] | **100.9**<br>98, [g] 87, [i]<br>84-320, [j]<br>99, [k] 103, [l]<br>102 [m] | **472.9**<br>461,[g] 583, [i]<br>224-440, [j]<br>443, [k] 468, [l]<br>444 [m] | **266.6**<br>259, [g] 271, [i]<br>232-250,[j]<br>344,[k] 269, [l]<br>240 [m] | **0.71** |
| **$ZrB_2$** | **540.5**<br>581, [n] 596, [o]<br>564-606 [p] | **55.9**<br>55, [n] 48, [o]<br>52-54 [p] | **111.1**<br>121, [n] 169, [o]<br>118-134 [p] | **422.2**<br>445, [n] 482, [o]<br>436-477 [p] | **252.7**<br>240, [m] 240, [o]<br>256-281 [p] | **0.78** |
| **$HfB_2$** | **592.7** | **99.6** | **141.3** | **481.3** | **262.3** | **0.81** |
| **$VB_2$** | **682.9**<br>699 [o] | **110.5**<br>146 [o] | **126.5**<br>109 [o] | **462.4**<br>552 [o] | **229.0**<br>167 [o] | **0.68** |
| **$NbB_2$** | **608.0**<br>517, [q] 290, [r] | **104.1**<br>95, [q] 110, [r] | **193.0**<br>120, [q] 190, [r] | **493.7**<br>528, [q] 474, [r] | **220.3**<br>112, [q] 225, [r] | **0.81** |
| **$TaB_2$** | **708.6** | **129.2** | **218.6** | **517.0** | **236.6** | **0.73** |
| **$CrB_2$** | **600.9** | **165.3** | **205.8** | **345.4** | **153.6** | **0.58** |
| **$MoB_2$** | **621.4** | **116.9** | **228.4** | **404.3** | **175.0** | **0.65** |
| **$WB_2$** | **636.1**<br>590-586 [s] | **130.0**<br>187-184 [s] | **263.4**<br>236-235 [s] | **380.9**<br>443-419 [s] | **137.1**<br>99-94 [s] | **0.60** |
| **$AgB_2$** | **220.9** | **253.7** | **82.5** | **173.5** | **-42.4** | ***u*** |
| **$AuB_2$** | **258.8** | **257.8** | **88.9** | **154.3** | **-49.4** | ***u*** |

* mechanically unstable phases, see the text.
[a]Reference 81, DFT (CRYSTAL98).
[b]Reference 54, TB-LMTO.
[c]Reference 63, FP-LAPW + GGA.
[d]Reference 82, DFT plane-wave method.
[e]Reference 83, DFT.
[f]Reference 84, LCAO-DFT.
[g]Reference 64, first-principles pseudopotential method (LDA-GGA)
[h]Reference 70, plane waves – DFT.
[i]Reference 67, Hartree-Fock.
[j]Experimenal data cited in Reference 67.
[k]Reference 85, FP-LAPW.
[l]Reference 86, DFT-CASTEP.

[m]Reference 69, CASTEP-GGA.
[n]Reference 28, experiment.
[o]Reference 16, LCAO-DFT.
[p]Reference 15, DFT (LDA-GGA).
[q]Reference 87, LCAO-DFT.
[r]Reference 88, CASTEP-GGA.
[s]Reference 76, CASTEP (LDA-GGA).

**TABLE IV**. Calculated bulk modules (*B*, in GPa) and shear modules (*G,* in GPa) for monocrystalline $AlB_2$-like diborides of *s, p* and *d* metals in comparison with available theoretical data.

| diboride | *B* | *G* | *B*/*G* |
|---|---|---|---|
| **$MgB_2$** | **156.7**<br>150, [a] 163, [b] 122-161, [c] 139, [d] 140,[e] | **131.5**<br>115, [a] 90, [d] | **1.19** |
| **$CaB_2$** | **122.2**<br>134, [a] | **111.5** | **1.10** |
| **$AlB_2$** | **196.5**<br>195- 211, [f] | **132.1** | **1.49** |
| **$ScB_2$** | **196.5** | **202.6** | **0.97** |
| **$YB_2$** | **156.7** | **150.0** | **1.05** |
| **$TiB_2$** | **260.7**<br>213, [a] 251-277, [d] 299-306, [g] 249-304, [h] | **270.6**<br>261, [d] 255, [h] | **0.96** |
| **$ZrB_2$** | **228.8**<br>195, [i] 272-275, [j] 237-260, [k] | **231.2** | **0.99** |
| **$HfB_2$** | **270.0** | **239.9** | **1.17** |
| **$VB_2$** | **283.9**<br>175, [i] 298, [j] | **246.5** | **1.18** |
| **$NbB_2$** | **298.9**<br>248, [l] 298, [m] | **219.8**<br>212, [m] | **1.22** |
| **$TaB_2$** | **340.8** | **243.8** | **1.36** |
| **$CrB_2$** | **300.1** | **170.7** | **1.55** |
| **$MoB_2$** | **310.5** | **192.0** | **1.62** |
| **$WB_2$** | **329.6**<br>322-327, [n] | **171.9**<br>141-144, [n] | **1.92** |

[a]Reference 54, TB-LMTO.
[b]Reference 84, LCAO-DFT.
[c] Reference 81, DFT (CRYSTAL98).
[d]Reference 64, first-principles pseudopotential method (LDA-GGA)
[e]Reference 89, FLAPW-GGA
[f]Reference 70, plane waves – DFT.
[g]Reference 67, Hartree-Fock.
[h]Reference 85, FP-LAPW.
[i]Reference 90, TB-LMTO.
[j]Reference 74, LCAO-DFT.
[k]Reference 73, DFT (LDA-GGA).
[l]Reference 87, LCAO-DFT.
[m]Reference 88, CASTEP-GGA.
[n]Reference 76, CASTEP (LDA-GGA).

**TABLE V**. Calculated values of some elastic parameters for $MB_2$ polycrystalline ceramics as obtained in the Voigt-Reuss-Hill approximation: bulk modules ($B_{VRH}$, in GPa), compressibility ($\beta$, in $GPa^{-1}$), shear modules ($G_{VRH}$, in GPa), Young's modules ($Y_{VRH}$, in GPa), Poisson's ratio (ν), Lamé's coefficients ($\lambda$, $\mu$, in GPa) and density (ρ, in $kg/m^3$) in comparison with available experiments.

| diboride | $B_{VRH}$ | $\beta$ | $G_{VRH}$ | $Y_{VRH}$ | ν | $\lambda$ | $\rho$ |
|---|---|---|---|---|---|---|---|
| **$MgB_2$** | **149.9**<br>145, [a] 120, [b] 151 [c] | **0.00667**<br>0.00690 [a] | **117.0**<br>117.5 [a] | **278.6** | **0.190** | **71.9** | **2695**<br>2650 [a] |
| **$CaB_2$** | **111.5** | **0.00897** | **88.9** | **210.7** | **0.185** | **52.2** | **2869** |
| **$AlB_2$** | **184.6**<br>170 [d] | **0.00541** | **98.2** | **250.2** | **0.274** | **119.3** | **3314**<br>3190 [e] |
| **$ScB_2$** | **198.9** | **0.00503** | **201.4** | **451.8**<br>480 [f] | **0.121** | **64.6** | **3749**<br>3670 [f] |
| **$YB_2$** | **171.5** | **0.00583** | **150.1** | **348.7** | **0.161** | **71.4** | **5155** |
| **$TiB_2$** | **263.2**<br>255, [h] 276, [i] 251, [j] | 0.00380<br>0.00392 [h]<br>0.00362, [i]<br>0.00398 [j] | **268.9**<br>252, [i]<br>237 [j] | **601.9**<br>372-551,[f]<br>579 [j] | **0.119**<br>0.110, [g]<br>0.108, [h]<br>0.151, [i]<br>0.141 [j] | **83.9** | **4591**<br>4520, [e]<br>4510, [i]<br>4510[10] |
| **$ZrB_2$** | **237.6**<br>218 [k] | **0.00421** | **230.5** | **522.6**<br>343-506 [e] | **0.134**<br>0.110 [e] | **84.0** | **6135**<br>6085, [e]<br>6100 [g] |
| **$HfB_2$** | **278.8**<br>222 [k] | **0.00359** | **239.5** | **558.6**<br>500 [e]<br>497 [k] | **0.166**<br>0.120 [g] | **119.2** | **11078**<br>11190 [e] |
| **$VB_2$** | **283.2** | **0.00353** | **244.1** | **568.8**<br>268 [e] | **0.165**<br>0.100 [e] | **120.5** | **5208**<br>5070 [e] |
| **$NbB_2$** | **327.3** | **0.00306** | **221.2** | **541.7**<br>637 [e] | **0.224** | **179.8** | **6952**<br>6970 [e] |
| **$TaB_2$** | **370.8** | **0.00270** | **244.0** | **600.3**<br>257 [e] | **0.230** | **208.2** | **12814**<br>12540 [e] |
| **$CrB_2$** | **315.7** | **0.00317** | **174.2** | **441.5**<br>221 [e] | **0.267** | **199.6** | **5533**<br>5160-5200 [e] |
| **$MoB_2$** | **369.4** | **0.00271** | **191.3** | **498.5** | **0.279** | **241.8** | **7533**<br>7780 [e] |
| **$WB_2$** | **436.8** | **0.00229** | **168.8** | **448.7** | **0.329** | **324.3** | **13150** |

[a]Reference 106.
[b]Reference 107.
[c] Reference 108.
[d]Reference 71.
[e]Reference 109.
[f]Reference 110.
[g]Reference 111.
[h]Reference 112.
[i]Reference 113.
[j]Reference 114..
[k]Reference 7.

# Figures

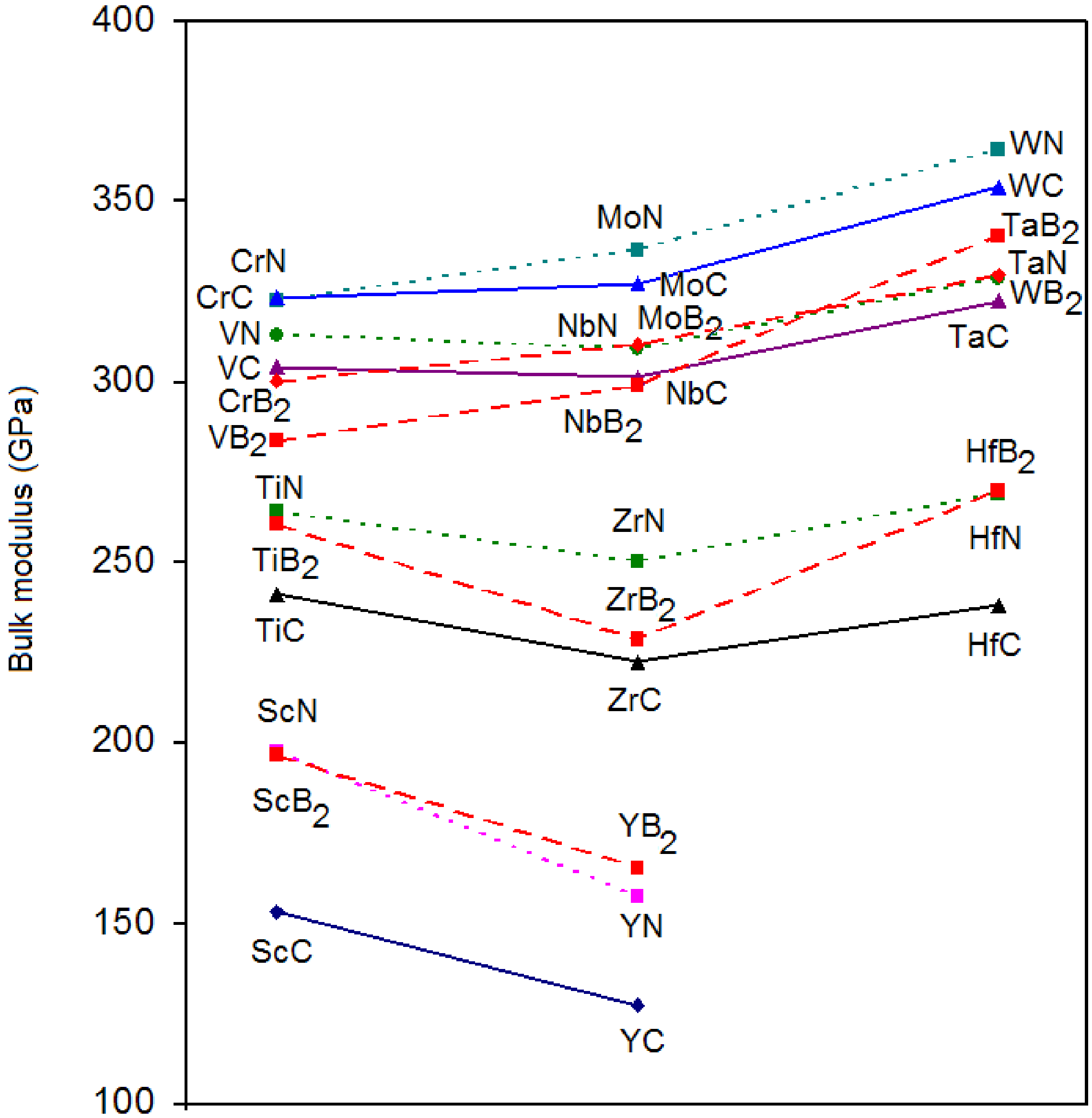

**FIG. 1** (*Color online)* The calculated bulk modules of monocrystalline $AlB_2$ –like diborides of 3*d* - 5*d* metals in comparison with bulk modules [95] of cubic nomocarbides and mononitrides of the same metals.

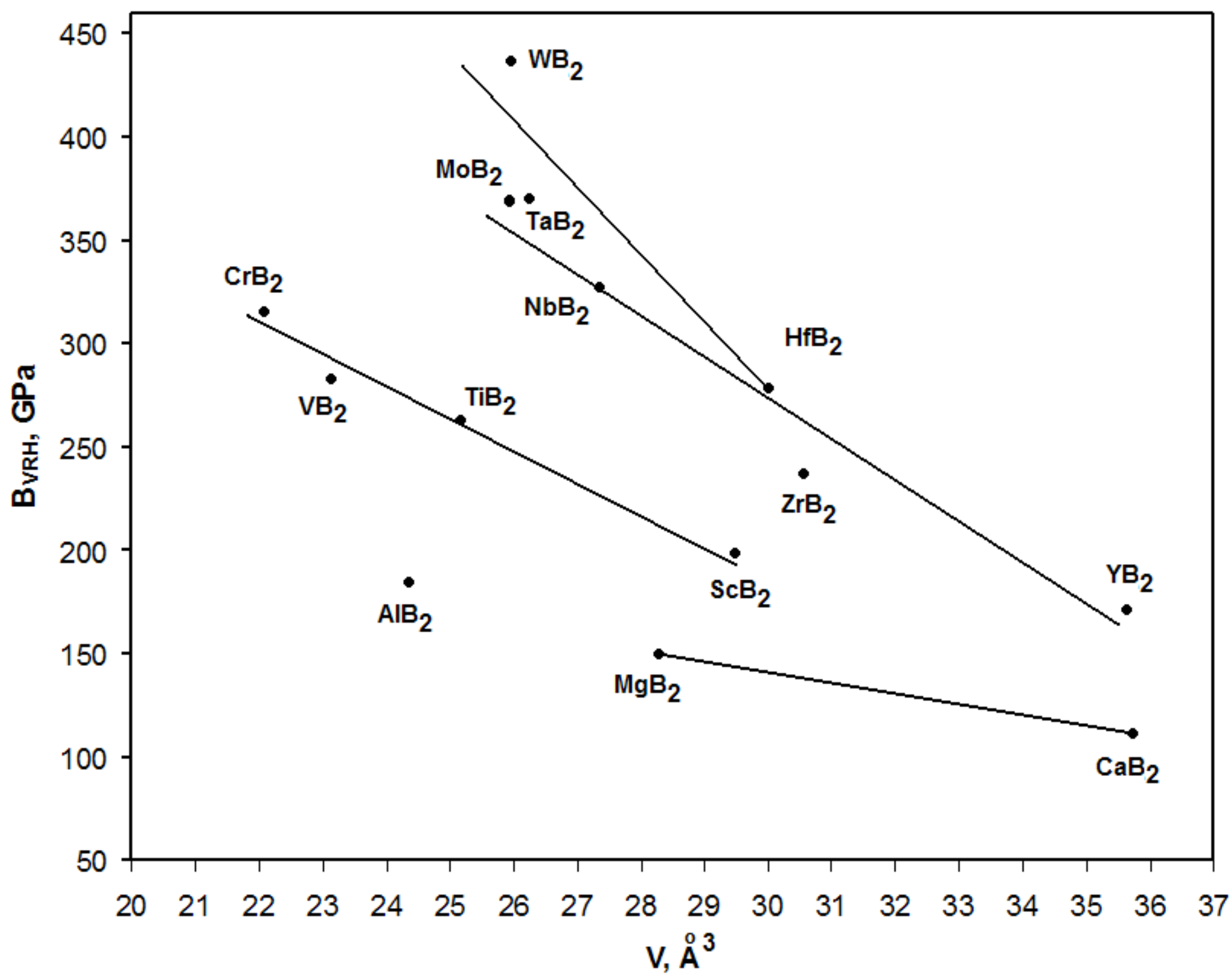

**FIG. 2**. The calculated bulk modules ($B_{VRH}$, in GPa) of polycrystalline $MB_2$ ceramics with respect to cell volumes of diborides.